\documentclass[usenatbib]{mn2e}
\usepackage{graphicx}

\title{CI Aquilae: a recurrent nova with an unusually long plateau phase}

\author[K. Matsumoto et al.]{K. Matsumoto$^{1}$\thanks{e-mail: 
       {\tt katsura@cc.okayama-u.ac.jp}}, R. Ishioka$^{2}$, 
       M. Uemura$^{2}$, T. Kato$^{2}$, T. Kawabata$^{3}$ \\
       $^{1}$ Graduate School of Natural Science and Technology,
              Okayama University, Okayama 700-8530, Japan\\
       $^{2}$ Department of Astronomy, Faculty of Science, Kyoto University, 
              Kyoto 606-8502, Japan\\
       $^{3}$ Bisei Astronomical Observatory, 1723-70 Ohkura, Bisei, 
              Okayama 714-1411, Japan
       }

\begin{document}

\date{Accepted.
      Received}
\pagerange{\pageref{firstpage}--\pageref{lastpage}}
\pubyear{}

\maketitle

\label{firstpage}

\begin{abstract}
We present the results of optical photometry of the recurrent nova 
CI Aql in later phase of the outburst which occurred in 2000.
Our observation revealed that the object reached the quiescent level 
between 2001 December and 2002 April and therefore that CI Aql is 
a unique recurrent nova characterized by an extremely long (1.4--1.7~yr) 
plateau phase.
The light curve obtained in the outburst suggests that the object 
is the first example of an intermediate between classical novae and 
recurrent novae.
In comparison with estimation given in published theoretical calculations, 
the long duration of the plateau phase supports a higher hydrogen 
content of the white-dwarf envelope, while such an abundance of 
hydrogen requires a later cessation of the wind which is in disagreement 
with the sudden fading observed in late November of 2000. 
The light curve obtained in later phase of the outburst indicates 
that the object was fainter and the gradual decline was steeper 
than predicted. 
These discrepancies between the observation and theoretical prediction 
require drastic modification of the present model of CI Aql.
\end{abstract}

\begin{keywords}
 accretion, accretion discs ---
 novae, cataclysmic variables ---
 stars: individual (CI Aql)
\end{keywords}

\section{Introduction}

Recurrent novae are one of subclasses of cataclysmic variables (CVs) 
which are binary systems containing a white dwarf as an accretor 
and a mass-donor star filling its Roche lobe. 
Non-magnetic CVs are generally classified into novae, novalike variables, 
and dwarf novae \citep[e.g.,][ for a review]{Warner1995_CV}. 
Novae are CVs showing outbursts due to thermonuclear runaway on 
the surface of the accreting white dwarf, and are classified into 
two types of classical novae and recurrent novae. 
The former are novae with single outburst detections, 
and a multiple record of outbursts gives a nova the latter classification. 
At present about 300 novae comprising about 30~\% of CVs are classified 
\citep{Downes2001_CVcat}\footnote{See also 
{\tt http://icarus.stsci.edu/\%7Edownes/cvcat/}.},
and only ten recurrent novae are known in them.

Differences between classical novae and recurrent novae principally 
result from a significant difference in mass of the accretor. 
White dwarfs in recurrent novae are believed to be massive 
compared with those in classical novae \citep[e.g.,][]{Starrfield1988_NR,
Kahabka1999_USco,Hachisu2000_USco-1,Thoroughgood2001_USco,Hachisu2001_NR}.
Such a condition produces frequent nova-explosions for recurrent novae 
occurring in shorter recurrence intervals of 10--100~yr compared with 
those for classical novae ($\ge10^{4}$~yr).
A recurrent nova generally shows a faster decline from its outburst, 
which mainly depends on a mass of the accretor and its envelope-mass 
at an eruption. 
In contrast, white dwarfs in classical novae are believed to have 
moderate masses and accumulate hydrogen-rich matter much slower 
than in recurrent novae. 
As a result, a part of the hydrogen diffuses into a white dwarf 
of a classical nova before an ignition so that a surface layer of 
the white dwarf is highly dredged up into the hydrogen-rich envelope 
and blown off in the outburst-wind \citep[e.g.,][]{Prialnik1986_N,Kato1994_N}.
This is consistent with spectra of ejecta observed in outbursts 
of classical novae which show heavy elements such as carbon, oxygen, 
and neon, and thus outbursts of a classical nova provide a gradual 
erosion for the white dwarf. 
On the other hand, ejecta observed in outbursts of recurrent novae 
are not enriched by such metals, i.e., massive white dwarfs in those 
systems are not eroded. 
Hence, in recurrent novae, the white dwarfs likely increase
mass toward the Chandrasekhar limit and are possibly fated 
to be type Ia supernovae \citep{Nomoto1984_SNIa,Nomoto1991_SNIa}.

The recurrent nova CI\,Aql was originally recorded as a possible nova 
in 1917 \citep{Reinmuth1925_CIAql}.
Lack of detailed information in that event had unfortunately left 
the nature of the object unsettled for a long time. 
The object is a peculiar eclipsing binary system showing 0.6~mag 
depth of primary eclipse with an orbital period of 0.618355~d 
\citep{Mennickent1995_CIAql}. 
The optical spectrum in quiescence shows higher-excited 
emission-features of He\,{\sc ii} and C\,{\sc iii}--N\,{\sc iii} 
complex on a reddened continuum, while all Balmer lines are detected 
as absorption lines \citep{Greiner1996_CIAql}. 
Such absorption features generally imply non-CV nature, and 
the original suspected-CV classification \citep{Duerbeck1987_Ncat}
had almost been disregarded. 
However, CI\,Aql underwent the second-recorded outburst in April of 
2000 \citep{Takamizawa2000_CIAql}, and spectroscopic observations 
revealed that the object is surely a recurrent nova
\citep[e.g.,][]{Uemura2000_CIAql,Kiss2001_CIAql,Burlak2001_CIAql}.
An optical light curve for the early part of the outburst was presented 
and discussed in \citet{Matsumoto2001_CIAql}.
We have made photometric observation of the object in 2001 and 2002 
which revealed an unexpected late evolution of the outburst.

\section{Observation and data reduction}

\begin{table}
\begin{center}
\caption{Log of the photometric observation. 
 The orbital phase coverages, represented by $\Phi$, 
 are based on the ephemeris given in \citet{Mennickent1995_CIAql}.}
\label{tab:log_phot}
\begin{tabular}{lrcccc}
\hline
\multicolumn{2}{c}{Date} & {HJD ($-$2450000.0)} & {error} & {$\Phi$} 
  & {Site$^{a}$} \\
\hline 
\multicolumn{2}{c}{(2001)} & & & & \\
Mar.\ & 15 & 1984.305--1984.317 &      & 0.86--0.87 & O \\
      & 19 & 1988.272--1988.343 & 0.03 & 0.26--0.38 & O \\
May   &  4 & 2034.274--2034.288 & 0.33 & 0.66--0.68 & K \\
      &  9 & 2039.286--2039.297 & 0.41 & 0.76--0.78 & K \\
      & 16 & 2046.290--2046.306 & 0.26 & 0.09--0.11 & K \\
      & 25 & 2055.265--2055.291 & 0.17 & 0.60--0.64 & K \\
      & 28 & 2058.255--2058.278 & 0.10 & 0.44--0.47 & K \\
      & 31 & 2061.275--2061.302 & 0.30 & 0.32--0.36 & K \\
Jun.\ &  7 & 2068.189--2068.200 & 0.08 & 0.50--0.52 & O \\
      &  8 & 2069.252--2069.265 & 0.17 & 0.22--0.24 & O \\
      & 12 & 2073.287--2073.291 & 0.11 & 0.75--0.75 & K \\
Jul.\ &  1 & 2092.282--2092.292 & 0.28 & 0.47--0.48 & K \\
      & 19 & 2110.216--2110.226 & 0.13 & 0.47--0.48 & K \\
      & 27 & 2118.214--2118.228 & 0.42 & 0.40--0.42 & K \\
      & 30 & 2121.156--2121.173 & 0.25 & 0.16--0.19 & K \\
Aug.\ &  4 & 2126.223--2126.227 & 0.58 & 0.35--0.36 & K \\
      & 11 & 2133.083--2133.097 & 0.09 & 0.45--0.47 & K \\
      & 14 & 2136.181--2136.196 & 0.11 & 0.46--0.48 & K \\
      & 23 & 2144.987--2144.996 & 0.11 & 0.70--0.71 & K \\
Sep.\ &  1 & 2154.010--2154.030 & 0.19 & 0.29--0.32 & K \\
      & 17 & 2170.094--2170.105 & 0.45 & 0.30--0.32 & K \\
Oct.\ & 11 & 2194.021--2194.029 & 0.39 & 0.00--0.01 & K \\
Dec.\ &  4 & 2247.885--2247.893 & 0.29 & 0.11--0.12 & K \\
\\
\multicolumn{2}{c}{(2002)} & & & & \\
Apr.\ &  5 & 2370.320--2370.331 & 0.30 & 0.11--0.12 & K \\
Jun.\ & 14 & 2440.150--2440.159 & 0.18 & 0.03--0.05 & B \\
Aug.\ & 26 & 2513.089--2513.117 & 0.10 & 0.98--0.02 & B \\
Oct.\ &  9 & 2556.933--2556.961 & 0.04 & 0.88--0.93 & B \\
\hline
\multicolumn{6}{l}{$^{a}$ O: Ouda, K: Kyoto, B: Bisei} \\
\end{tabular}
\end{center}
\end{table}

\begin{table}
\begin{center}
 \caption{A multi-color magnitude of CI\,Aql on 2001 March 15, 
 which was obtained within a duration of orbital phase of 0.86--0.87.}
\label{tab:multi}
\begin{tabular}{lrr}
\hline
{}          & {mag.} & {error} \\
\hline
$B$         & 15.80 & 0.15 \\
$V$         & 14.81 & 0.05 \\
$R_{\rm c}$ & 14.24 & 0.01 \\
$I_{\rm c}$ & 13.53 & 0.04 \\
\hline
\end{tabular}
\end{center}
\end{table}

The photometric observations were conducted on 27 nights 
between 2001 March 15 and 2002 October 9 at three sites of 
the Ouda station and the rooftop of Department of Astronomy,
Kyoto University, and the Bisei Astronomical Observatory 
(Table~\ref{tab:log_phot}).

The Ouda observations were made by using an SITe SI004AB CCD chip 
(PixelVision) attached to a Ritchey-Chretien Cassegrain telescope 
with a 60-cm aperture. 
An $R_{\rm c}$ filter was applied, and we made multi-color photometry 
in $B, V, R_{\rm c}$, and $I_{\rm c}$-bands on March 15 
(Table~\ref{tab:multi}).
Exposure time was set to 30~s on March 15, June 7, and 8.
On March 19, exposure time was varied to 15--60~s depending on 
sky conditions. 
The data reduction and analysis for which we used aperture photometry 
were performed by using the {\sc iraf}\footnote{{\sc iraf} is 
distributed by the National Optical Astronomy Observatories, which are 
operated by the Association of Universities for Research in Astronomy, 
Inc., under cooperative agreement with the National Science Foundation. 
See {\tt http://iraf.noao.edu/} for more information.}.

The Kyoto observations were made by using SBIG ST-7 and ST-7E 
CCD cameras attached to Schmidt-Cassegrain telescopes with 25 and 
30-cm apertures (Meade LX-200).
No filter was applied, and exposure time was set to 30~s. 
The data reduction and analysis for which we used PSF photometry 
were performed by a Java$^{\rm TM}$-based aperture and PSF 
photometry-package developed by one of the authors (T.~Kato).

The Bisei observation was made by using Mutoh CV-16II and CV-16IIE 
CCD cameras attached to a classical Cassegrain telescope with a 101-cm 
aperture.
An $R_{\rm c}$ filter was applied, and exposure time was set to 30~s.
The data reduction and analysis for which we used aperture photometry 
were performed by using the {\sc iraf}.

The brightness of the object was determined relatively to 
a local comparison star and a check star that confirms constancy 
of the comparison star.
Those stars were GSC\,5114.149 and GSC\,5114.584 in the Ouda and Kyoto 
observations, and USNO\,0825.13270569 and GSC\,5114.247 in the Bisei 
observation. 
We used A.~Henden's photometric sequence for the observed 
field.\footnote{\tt ftp://ftp.nofs.navy.mil/pub/outgoing/aah/sequence/CIAql.dat}
Magnitudes obtained in the Kyoto observations are corresponding to ones 
in $R_{\rm c}$, according to sensitivities of the CCD chips. 

\begin{figure}
 \resizebox{\hsize}{!}{\includegraphics[angle=270]{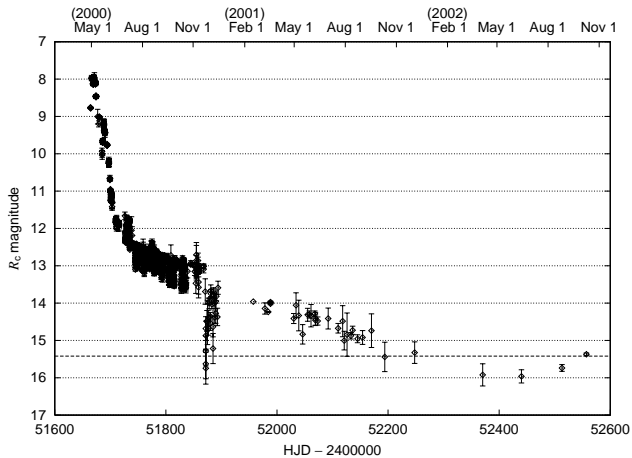}}
 \caption{The optical light curve of the 2000 outburst of CI\,Aql is shown. 
 The horizontal dashed line represents the expected quiescent level 
 at outside of the eclipse.}
 \label{fig:lc_all}
\end{figure}

\section{Remarkably long plateau stage}

Fig.~\ref{fig:lc_all} shows an optical light curve of CI\,Aql 
for two years after the 2nd recorded eruption in 2000. 
The light curve consists of the early part of the outburst 
described in \citet{Matsumoto2001_CIAql} and the present results 
given in Table~\ref{tab:log_phot} and \ref{tab:multi} ($R_{\rm c}$ only).
It is noted that the data on February 16, March 9, and April 30 of 2001 
are excluded from Table~\ref{tab:log_phot}, because they were already 
presented in \citet{Matsumoto2001_CIAql}.

The light curve indicates that the object kept a gradual decline 
of about $1~{\rm mag}/200$~d (plateau phase) since a sudden fading 
and recovery observed in late November of 2000 (around JD 2451870) 
after which the object became systematically about 0.5~mag fainter.
On 2001 May 16, the object was observed to be somewhat fainter 
in the gradual decline, which was very likely due to a primary eclipse 
of the binary system (Table~\ref{tab:log_phot}).

\citet{Szkody1994_Novae} reported $m_{V}=16.22$ and $V-R=0.68$ 
for the object in the previous quiescence.
This estimation is in agreement with an independent record made by 
the RoboScope between 1995--1996 which shows a $m_{V}=$16.1--16.7 
orbital variation with typical accuracy of about 0.1~mag 
(K.~Honeycutt, private 
communication\footnote{It has been pointed out that the photometric 
$V$-magnitudes in 1991--1995 presented in fig.~1 of 
\citet{Mennickent1995_CIAql} is insecure due to an unknown 
zero-point in their photometric calibration.}).
Thus, in the quiescence it is expected that CI\,Aql is 
$m_{R}\sim15.4$ and $\sim16.0$~mag at outside and minimum of 
a primary eclipse, respectively, with an assumptively constant 
$V-R=0.68$ obtained at outside of an eclipse. 
The validity of that quiescent level is also supported by 
identifications for the object in records of POSS~II 
(GSC2~S300213319926: $m_{F}=15.79$, $m_{V}=16.05$) and USNO 
(USNO~0825.13277160: $m_{\rm red}=15.7$, $m_{\rm blue}=16.8$). 
As seen in Fig.~\ref{fig:lc_all}, the object was significantly 
brighter than the quiescent level in 2001.

The observation on 2001 October 11 indicates that the object 
was about 15.44~mag which was comparable to the quiescent level.
The orbital phase of the binary system was 0.0--0.01 
at the observation, i.e., the timing corresponded to a minimum 
of the primary eclipse.
Hence, assuming a 0.5--0.6~mag depth of the primary eclipse, 
the brightness observed on October 11 was consistent 
with the declining trendline, and we can conclude that 
CI\,Aql had still not reached to quiescence at around JD 2452200. 
The next data-point on 2001 December 4 again demonstrates 
the same state of the object: the object was brighter than 
the quiescent level while the orbital phase of the object 
was 0.11--0.12, i.e., in the middle of the egress.

In the observation on 2002 April 5, however, we found that 
the object was 15.92~mag which was not only considerably fainter 
than in 2001 but also $\sim0.4$~mag fainter as compared with 
the expected quiescent level. 
The orbital phase was 0.11--0.12 at the observation, and 
a magnitude expected for outside of the primary eclipse 
on that occasion is well consistent with the quiescent level.
Thus, we conclude that CI\,Aql was not in quiescence at least 
prior to 2001 December 4, and reached the quiescence prior to 
2002 April 5, after the plateau phase which continued for 
an acceptable range of 510--630~d. 
This result is well consistent with the result given in 
\citet{Lederle2002_CIAql}, and our observations on 2002 June 14, 
August 26, and October 9 confirm the conclusion.

\section{Possible new class of recurrent novae}

\subsection{Distinctive properties}

Most recurrent novae except T\,Pyx had been known as very fast novae 
which decline with $t_{3}\sim10$~d and completely fade to quiescence 
within several months in nova-outbursts.
T\,Pyx is an exceptional recurrent nova with the shortest 
orbital period of 1.8~h \citep{Schaefer1992_TPyx}, 
and it probably consists a dwarf companion as the mass donor.
The rest of recurrent novae are divided into two subclasses of 
the RS\,Oph-type and U\,Sco-type which probably contain a red giant 
and a slightly evolved main-sequence star as mass donors, 
respectively \citep{Hachisu2001_NR}.
CI\,Aql apparently belongs to the latter based on the orbital period, 
and the extremely long decline of $t_{3}\sim35$~d
makes the object a peculiar member of the subclass  
\citep{Matsumoto2001_CIAql}.
Such a slower decline suggests a smaller mass of the white dwarf 
of CI\,Aql compared with those of other recurrent novae. 
This nature is also supported by the long duration of the plateau 
phase and the longest recurrence interval of 83~yr in recurrent novae. 
Similar characteristics have been found in a recurrent nova IM\,Nor 
in the 2002 outburst: a slow decline of $t_{3}\sim50$~d and 
a recurrence interval of 82~yr are comparable to those of 
CI\,Aql\footnote{Since significant variations were superimposed 
on the decline of the 2000 outburst of CI\,Aql, $t_{3}$ of the object 
is possibly larger than the value described above (see fig.~3 of 
\cite{Kato2002_IMNor}).} \citep{Kato2002_IMNor}.
All recurrence intervals in recurrent novae except CI\,Aql and IM\,Nor 
are a few decades, which makes those novae easier to be detected 
as recurrent objects, and therefore most of them are suspected to 
contain extremely massive white dwarfs near the Chandrasekhar limit 
\citep[e.g.,][]{Hachisu2000_USco-1,Hachisu2000_V394CrA,Hachisu2001_NR}.
Outbursts of classical novae are observed at only once for each, 
since intervals of $\ge10^{4}$~yr for outbursts are enough longer 
than the human history. 
Hence CI\,Aql is probably the recurrent nova closest to classical novae 
we have known.

The recorded amplitude of the 2000 outburst of CI\,Aql is 
approximately 7.5~mag as depicted in the light curve, assuming 
that the maximum occurred on 2000 May 1.
This amplitude is reasonable for an outburst of a recurrent nova 
($\sim$8--11~mag), and is significantly smaller than that of 
a classical nova (typically 9--14~mag).
However, as pointed out in \citet{Matsumoto2001_CIAql}, 
the firm occasion of the maximum is possibly uncertain, 
i.e., a brighter maximum which had occurred between 2000 April 11 
and 30 is still possible. 
In recurrent novae a dispersion of about 3~mag is seen in the distribution 
of amplitudes of outbursts, which can be interpreted as an effect of 
a correlation between absolute magnitudes and inclination angles for 
novae \citep{Warner1986_N}. 
Larger outbursts with amplitudes of $\ge$10~mag are observed 
in V394\,CrA, IM\,Nor, U\,Sco, and V745\,Sco, and these objects 
are known or suspected as binary systems with higher inclination 
angles in recurrent novae 
\citep[e.g.,][]{Schaefer1990_NR,Hachisu2000_V394CrA,
Hachisu2000_USco-1,Hachisu2001_NR,Kato2002_IMNor}.
The orbital light curve of CI\,Aql showing deep eclipses strongly 
suggests a higher inclination angle, which permits a possibility 
of an about 3~mag brighter maximum for the 2000 outburst of CI\,Aql; 
otherwise the object is a recurrent nova with an exceptionally smaller 
amplitude of an outburst. 
If the maximum indeed occurred at about the middle of 2000 April 
and if the 1917 outburst had been missed, the 2000 outburst of 
CI\,Aql might have been indistinguishable from a typical outburst 
of a classical nova.

In conclusion, these characteristics of CI\,Aql observed in the 2000 
outburst indicate that this object is a recurrent nova having 
resemblances to classical novae in many aspects. 
We have found a recurrent nova showing a significantly slower 
evolution of an outburst which is likely attributed to 
a smaller mass of the white dwarf.
CI\,Aql is a very suggestive system of the first case of 
an intermediate between classical novae and recurrent novae.

\subsection{Incompatibility with the current model}

The 2000 outburst of CI\,Aql was theoretically modeled in 
\citet{Hachisu2001_CIAql-1,Hachisu2002_CIAql-2} 
by means of light curve analyses. 
Although the calculations well reproduced the observed light curve 
in the early part of the outburst prior to the plateau phase, 
the reproduction of the later phase suffered from an uncertainty 
in the duration of the plateau phase.

A principal discrepancy is that the observed duration of 
the plateau phase is evidently longer than the expectations. 
The light curves for the early parts of the outbursts in 1917 
and 2000, providing decline-rates of the fadings from 
the eruptions \citep{Williams2000_CIAql,Matsumoto2001_CIAql}, 
seem to tightly constrain a mass of the white dwarf to 
approximately $1.2~{\rm M_{\odot}}$.
This is especially clear in the 1917 outburst which was reproduced 
with a hydrogen content of $X=0.7$ \citep{Hachisu2001_CIAql-1}. 
As for the latter part of the 2000 outburst, the solution with 
$X=0.70$ is apparently suitable for the observed duration of 
the plateau phase. 
Hence a preferable situation is likely such a higher hydrogen 
content in the white-dwarf envelope of CI\,Aql.
In this case, a major problem should be solved is a discrepancy 
on the occasion of the wind-stop which was observed in late 
November of 2001 as a drop and prompt recovery of the brightness.
That sudden event is the most significant decline seen in the plateau 
phase and is therefore plausibly attributed to the wind-stop, which 
is the reason why helium-enriched cases for the white-dwarf envelope 
were considered in \citet{Hachisu2001_CIAql-1,Hachisu2002_CIAql-2}, 
and no other explanation for the event has been proposed at present.

In the plateau phase, the object was predicted to exhibit 
a supersoft X-ray emissivity as a consequence of hydrogen burning 
on the surface of the white dwarf
\citep{Hachisu2001_CIAql-1,Hachisu2002_CIAql-2}. 
In the 1999 outburst of a recurrent nova U\,Sco, supersoft X-ray was 
detected from that object in the plateau phase \citep{Kahabka1999_USco}.
The supersoft X-ray phase had been predicted by a theoretical model 
of hydrogen-burning surfaces of white dwarfs for recurrent novae 
\citep{Kato1996_SSS}. 
\citet{Matsumoto2002_USco} obtained a photometric evolution 
of the 1999 outburst, and an orbital-period change from which 
the mass-transfer rate during the previous quiescence between 1987 
and 1999 was observationally estimated was detected in the light curve.
These observations of U\,Sco were successfully explained by 
a consistent model based on the scheme for recurrent novae
\citep{Hachisu2000_USco-1}, and the physical parameters led 
U\,Sco to the most probable candidate of a progenitor system 
of a type Ia supernova. 

The expected mass of the white dwarf of CI\,Aql is a reasonable 
value for a supersoft X-ray source \citep[e.g.,][]{vdH1992_SSS}, 
though a definite detection of supersoft X-ray has not been reported 
in the 2000 outburst.
Such softer X-ray may be difficult to be detected if it exists, 
because of the higher interstellar reddening on the line of sight 
toward the object which is demonstrated by several optical 
observations \citep{Mennickent1995_CIAql,Greiner1996_CIAql,
Kiss2001_CIAql,Burlak2001_CIAql}.
The prediction for the supersoft X-ray phase is also supported 
by a similarity in the shape of the orbital modulation during 
the plateau phase compared with shapes observed in orbital light curves 
of known supersoft X-ray sources with higher inclination 
angles \citep{Matsumoto2001_CIAql}.

At the end of the plateau phase, a termination of the hydrogen 
burning should involve a final fading after which the outburst 
completely ended. 
In Fig.~\ref{fig:lc_all} we can see that no significant change of 
the plateau state was observed, implying that the hydrogen burning 
likely continued at least until the beginning of 2002.
As described above, if the occasion of the wind-stop is fixed, 
a longer duration of the supersoft X-ray phase is required 
to interpret the observed epoch for the end of the outburst.

Other discrepancies between the observation and calculation are 
the brightness and decline-rate of the object in the plateau phase. 
The former was inevitably caused by lack of secure observation 
published to be compared with calculations, as described in 
the footnote of section~3, i.e., the reproduced brightness of 
the object is brighter than the observation in the plateau phase 
(e.g., the multi-color magnitudes on 2001 March 15 provide 
the differences).
The decline-rate observed in the plateau phase was about 1~mag$/200$~d.
In contrast, the gradient of the reproduced light curve is too flat 
to fit the observed light curve for the case of $X=0.7$ 
which is suitable to explain the duration of the plateau phase. 
A revised model in \citet{Hachisu2002_CIAql-2} showed 
a better agreement for the decline-rate, but it requires 
a termination of the hydrogen burning to reduce the brightness, 
which necessarily shortens the duration of the plateau phase 
in contradiction with the observation. 

In conclusion, the duration of the plateau phase of the 2000 
outburst suggests a higher hydrogen content of the white-dwarf 
envelope for CI\,Aql based on the current model of recurrent novae.
However, the detailed behaviour observed in the later phase of 
the present outburst is inconsistent with the published models, 
which requires drastic modification of the present model.

\section*{Acknowledgments}
KM thanks K.~Ayani and M.~Ioroi for their supports 
during the observation at Bisei Astronomical Observatory.


\bsp
\label{lastpage}
\end{document}